\documentclass[letterpaper, 10 pt, conference]{ieeeconf}  

\IEEEoverridecommandlockouts                              

\usepackage{graphicx}
\usepackage{url}  
\usepackage{pifont}
\usepackage{amsmath}

\usepackage{lipsum}

\title{\LARGE \bf
Aud-Sur: An Audio Analyzer Assistant for \\ Audio Surveillance Applications
}

\author{
Phat~Lam$^{*}$,
        Lam~Pham$^{*}$, 
        Dat~Tran,
        Alexander~Schindler, \\
        Silvia~Poletti, 
        Marcel~Hasenbalg, 
        David Fischinger,
        Martin~Boyer       
\thanks{L. Pham, A. Schindler, S. Poletti, H. Marcel, F. David, 
and M. Boyer are with Austrian Institute of Technology, Vienna, Austria} 
\thanks{P. Lam is with Ho Chi Minh University of Technology, Vietnam}
\thanks{D. Tran is with FPT University, Vietnam} 
\thanks{(*) Main and equal contribution into the paper.}
}

\begin{document}

\maketitle
\thispagestyle{empty}
\pagestyle{empty}

\begin{abstract}
In this paper, we present an audio analyzer assistant tool designed for a wide range of audio-based surveillance applications (This work is a part of our DEFAME FAKES and EUCINF projects).
The proposed tool, refered to as Aud-Sur, comprises two main phases Audio Analysis and Audio Retrieval, respectively. 
In the first phase, multiple open-source audio models are leveraged to extract information from input audio recording uploaded by a user.
In the second phase, users interact with the Aud-Sur tool via a natural question-and-answer manner, powered by a large language model (LLM), to retrieve the information extracted from the processed audio file.
The Aud-Sur tool was deployed using Docker on a microservices-based architecture design. 
By leveraging open-source audio models for information extraction, LLM for audio information retrieval, and a microservices-based deployment approach, the proposed Aud-Sur tool offers a highly extensible and adaptable framework that can integrate more audio tasks, and be widely shared within the audio community for further development.

\indent \textit{Items}--- audio, surveillance, pre-trained model, large language model, retrieval.
\end{abstract}

\section{Introduction}
\label{intro}

With advancements in neural network architectures and deep learning techniques, various audio-based analysis tasks have been proposed by the research community such as audio captioning, audio event detection, speech-to-text, speech emotion detection, etc.  
However, these tasks are often proposed for specific purposes and evaluated on certain benchmark datasets.
For instance, audio captioning, audio event detection, speech to text, speech emotion, etc. have been evaluated using the large benchmark datasets such as Clotho~\cite{aud_cap_dataset}, Audioset~\cite{audioset}, Librispeech~\cite{lib_dataset}, and IMOCAP~\cite{imocap_dataset}, respectively.
Recently, some pretrained audio models leveraging self-supervised learning strategies have been proposed and proved potential across different audio tasks~\cite{whisper, wav2vec20, wavlm}.
However, these models are only verified and suitable for certain relevant audio tasks (i.e., mainly for human speech analysis), whereas there is a wide range of audio processing tasks requiring different audio engineering features.
For example, while the Mel filter bank is suitable for speech-to-text, linear filters are more well-suited for deepfake speech detection~\cite{survey_01}.
In Acoustic Scene Classification (ASC), a combination of multiple spectrograms of Gammatonegram (GAM), Constant-Q Transform (CQT), and Mel-spectrogram (MEL) has been shown to enhance model performance~\cite{lam_002}.
Furthermore, these pre-trained audio models present the high computational complexity and require the fine-tuning process for downstream audio tasks.
\begin{figure}[t]
    \centering
    \includegraphics[width=0.9\linewidth]{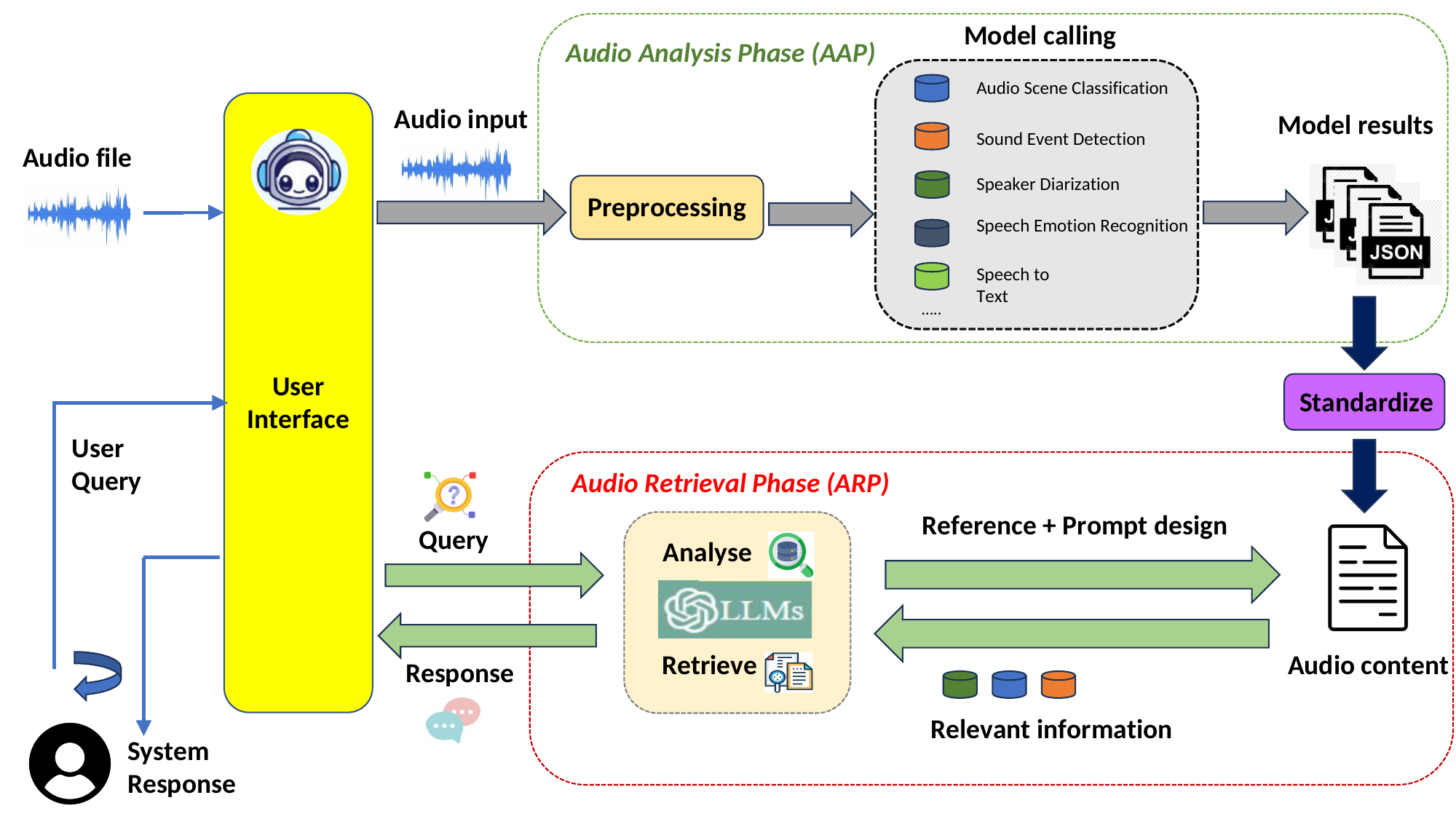}
    \caption{The high-level architecture of our proposed Aud-Sur tool}
    \label{fig:overall}
\end{figure}

In real-life audio-based applications, multiple audio tasks have to be integrated into complex systems rather than relying on a standalone task.
For example, a call center support tool may need to integrate speech denoise, eco reduction, speech-to-text, speaker identification, etc.
Therefore, the ability to seamlessly integrate new audio tasks is significantly essential for real-world audio-based applications. 
In the context of audio surveillance, these applications focus on audio event detection and audio tracking~\cite{app03}.
Existing audio surveillance systems are often designed for specific tasks such as anomaly violent detection~\cite{app01}, children abuse by voice~\cite{app02}, or riot context detection~\cite{lam_002, lampham-cbmi}. However, this task-specific approach could limit scalability
and the ability to integrate or extend new audio tasks in the future.
Furthermore, publications for audio surveillance often focus on model developments rather than proposing a deployment process that leads to a lack of comprehensive solution~\cite{survey_p}.
\begin{figure*}
    \centering
    \includegraphics[width=0.9\linewidth]{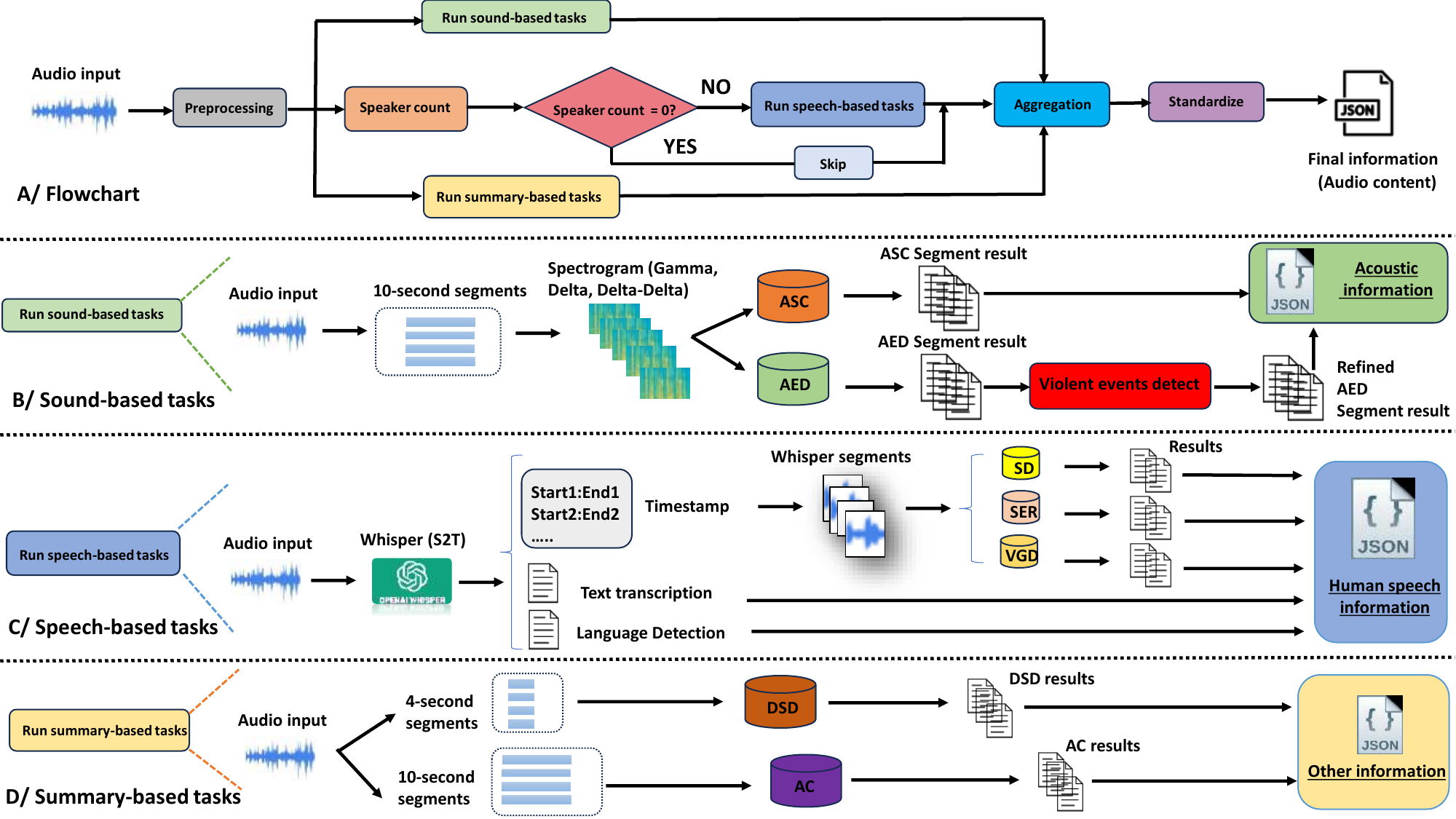}
    \caption{Logical processing flow in the Audio Analysis Phase (AAP)}
    \label{fig:audio_logic}
\end{figure*}
In this paper, we propose an audio analyzer assistant tool designed for a wide range of audio-based surveillance applications. 
We address the challenges from both audio research and audio application communities recently mentioned, then introducing following solutions:
\begin{itemize}
\item We leverage state-of-the-art open-source audio models, each excel in a specific task. This approach allows seamless upgrades, where an outdated model can be replaced with a more advanced one without requiring major system modifications.
\item Unlike conventional audio surveillance systems that focus only on audio event/scene detection, our tool integrates a broader range of tasks such as speech-to-text, speech emotion detection, audio captioning, language detection to provide richer context and deeper situational awareness.
\item We use a large language model (LLM) to enable natural and human-like interaction between our proposed system and user. This enhances accessibility and makes the tool more user-friendly both technical and non-technical users.



\item To enhance scalability and adaptability, we design the proposed system using a microservices-based architecture and deploy it using Docker. The modular design allows seamless updates, parallel execution, and easy integration of new models without modifying the entire system. 
\end{itemize}
\section{High-Level Architecture}
\label{architecture}

Figure~\ref{fig:overall} presents the high-level architecture of the proposed Aud-Sur tool. In overall, the tool comprises three main components. The first component is the User Interface (UI).
The UI serves as an entry point where users interact with the tool by uploading an audio file and submitting a query or question, typically in natural language, to request specific information about the audio. 
The second component is the Audio Analysis Phase (AAP) which processes the uploaded audio recording by performing various audio tasks. The AAP component extracts task-structured results representing multiple aspects of the audio input. 
For each audio task, the task's result is exported in a JSON file, which is then aggregated and standardized to ensure consistency across factors such as time boundaries, missing values among JSON results from other audio tasks.
Finally, the Audio Retrieval Phase (ARP)
utilizes the standardized results to interpret the extracted information and retrieve relevant insights based on the user's question, formulating a corresponding response in natural language.

\subsection{Audio Analysis Phase (AAP)}
The Audio Analysis Phase (AAP) aims to extract structured information representing different aspects of the audio recording.
The extracted data serves as the main reference resource for generating responses in the Audio Retrieval Phase (ARP). 
Since the quality of extracted information directly impacts the accuracy and relevance of the final response, ensuring high-quality results in this phase is crucial. 
Therefore, instead of relying on a single model with multiple tasks, we employ multiple individual models, each of which was optimized for a specific audio task and achieved high performance on benchmark datasets. 

To describe the AAP in detail, we introduce the Fig.~\ref{fig:audio_logic} in which a flow chart to process input audio is comprehensively presented.
In particular, at the top of the flow chart (Fig.~\ref{fig:audio_logic}a), the APP process begins with the Preprocessing module, which performs multiple steps such as resampling, filtering, and changing representation format (e.g., raw audio, spectrogram, MFCC) to ensure the alignment between the input audio and each of pre-defined single models. 
Then the preprocessed audio is fed into multiple audio models to extract audio information.
To assure a more organized and effective extraction of relevant information from an audio recording, we categorize multiple audio models in this phase into three main groups, referred to as the sound-based tasks, speech-based tasks, and summary-based tasks.
These three task groups analyze the input in parallel. 
Notably, for speech-based tasks, a preliminary speaker count module~\cite{pyannote} is first leveraged to detect the presence of human speech. If at least one speaker is identified, the system continues speech-based tasks processing; otherwise, these tasks are skipped to optimize computational efficiency. 
\begin{figure}[t]
    \centering
    \includegraphics[width=0.95\linewidth]{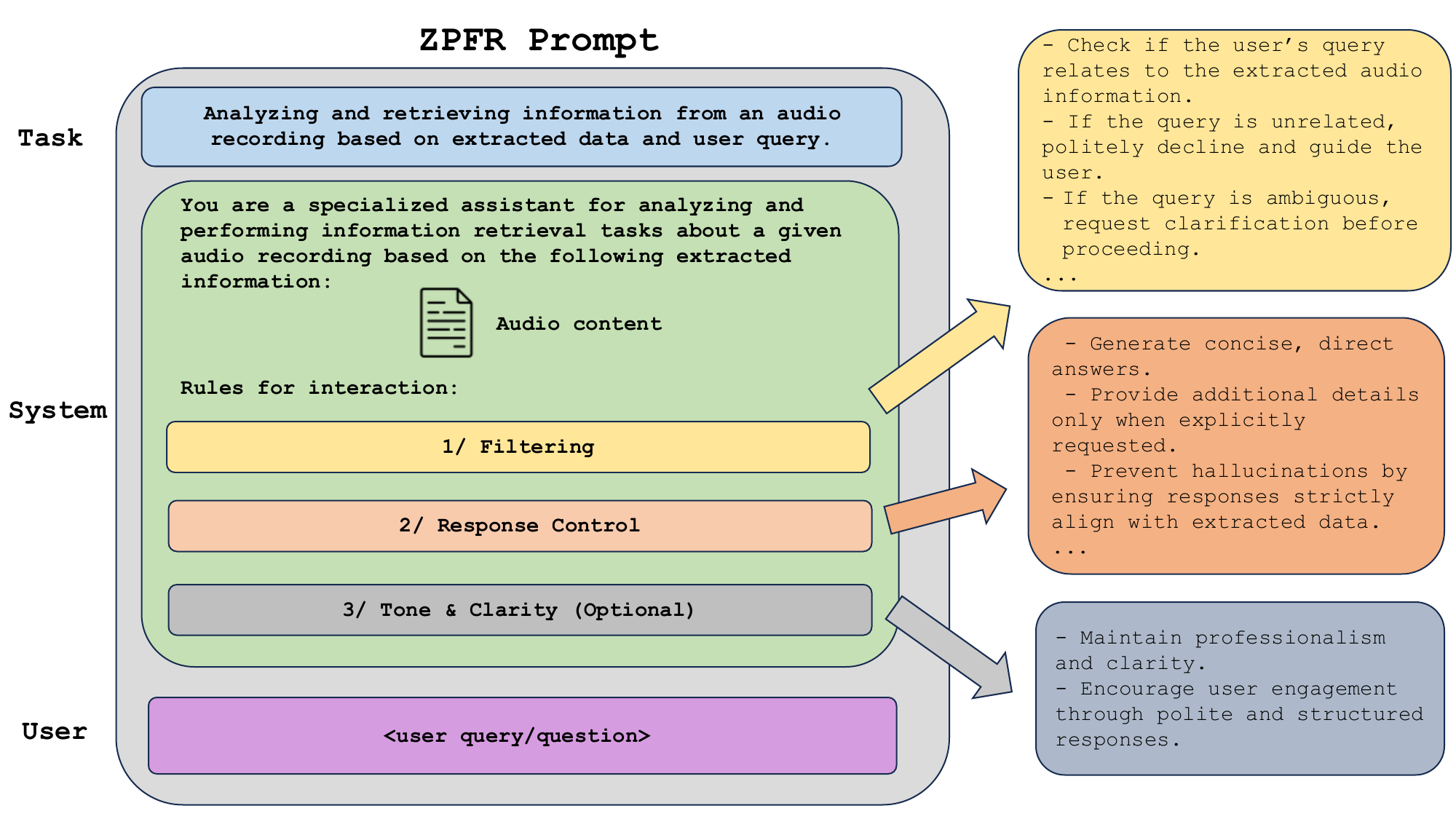}
    \caption{Zero-shot Prompting with Filtering \& Response control (ZPFR)}
    \label{fig:zpfr}
\end{figure}

For each task group, the detailed process is presented in Fig.~\ref{fig:audio_logic}b, Fig.~\ref{fig:audio_logic}c, and Fig.~\ref{fig:audio_logic}d, respectively.
The first group, referred to as the sound-based tasks, focuses on analyzing the background environment and non-speech elements within the recording. 
They include specific audio tasks of Acoustic Scene Classification (ASC)~\cite{lampham_01,lampham_02} (e.g., recordings into one of multiple pre-defined acoustic scene classes such as `airport', `park', `outdoor', `in door', etc.), Acoustic Event Detection (AED)~\cite{kong_pretrain} (e.g., recognizes specific auditory events such as `sirens', `alarms', or `footsteps', etc). 
As the tool is applied for surveillance applications, we highly focus on detecting events indicating violent contexts such as `gun', `fireworks', `screaming', `explosion', etc, by adding violent/alarm tags on the task results for the next ARP processing. As Fig.~\ref{fig:audio_logic}b describes, we employ sound-based tasks using 10-second segments instead of input recording for greater flexibility in handling variations in background environments and sound events. Within this task group, both the Acoustic Scene Classification (ASC) model and the Audio Event Detection (AED) model process image-like inputs, which are constructed from concatenating three types of features: spectrogram with Gammatone filter, delta (differential) and delta-delta (acceleration) coefficients derived from MFCC feature. Then, the segment-based results of these two models are gathered and 
collectively referred to as 'Acoustic information'. Notably, before finalizing the AED results, we apply an additional refinement step, in which we assign a `violent' tag to any segments containing at least one event indicative of violent behaviors. Among the 500+ class labels in AudioSet~\cite{audioset}, which the AED model was trained on, we manually selected a subset of class events associated with violence or riots and used them as label references for this tagging process. This serves as additional auditory cues in the next ARP to better 
enhance situational awareness for the main purpose of audio surveillance.

The second group, referred to as speech-based tasks, is designed to extract information related to human speech. 
These incorporate Speech-to-text (S2T)~\cite{whisper} (e.g., transcribing spoken content into text), Speaker Diarization (SD)~\cite{speechbrain} (e.g., segmenting audio based on personal identity or simply identifies `who spoke when'), Speech Emotion Recognition~\cite{speechbrain, wav2vec2} (e.g., determining the speaker’s emotional state) and Language Detection (LID)~\cite{whisper} (e.g., detecting the spoken language). The flow to process results in this task group is described in Fig.~\ref{fig:audio_logic}c. Firstly, a Speech-to-Text model (e.g., Whisper) is utilized to generate text transcriptions, which are segmented based on their respective time boundaries. Since Whisper is a general-purpose model, we derive language identification results directly from the language tag in its multitask sequence tokens. Using the time boundaries provided by Speech-to-text model, we further divide the audio input into segments, which then serve as inputs for the three following models: Speech Emotion Recognition (SER), Speaker Diarization (SD), and Voice Gender Detection (VGD). In other words, these models process audio segments defined by Whisper’s segmentation. Assuming Whisper’s time boundary estimation is sufficiently accurate, this approach of segmentation ensures that speaker-related attributes such as identity, gender, and emotion are captured precisely at the moment of speech, reducing errors associated with multiple speakers overlapping in a single segment when using fixed-length segments that could affect the performances of the next three single models. The results of single models in this task group are also aggregated, referred to as `Human speech information'.

\begin{figure}[t]
    \centering
    \includegraphics[width=0.95\linewidth]{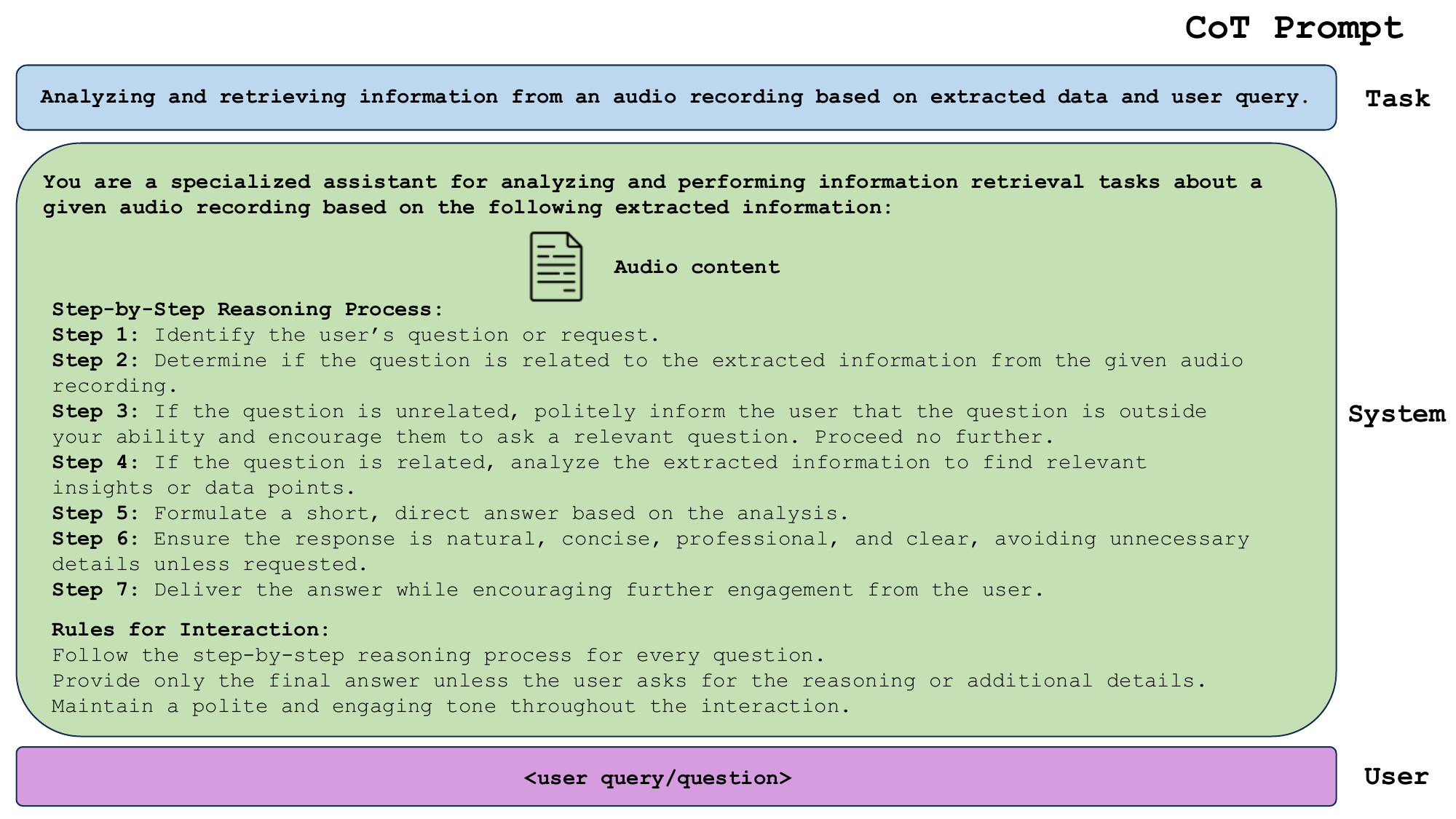}
    \caption{Chain-of-thought (CoT) prompting}
    \label{fig:cot}
\end{figure}
\begin{table}[t]
	\caption{Individual audio models integrated in the AAP component} 
	\vspace{-0.1cm}
	\centering
    \scalebox{0.8}{
	\begin{tabular}{|c |c|c| } 
		\hline 
		\textbf{Models}   &\textbf{Tasks} &\textbf{Licenses}  \\ 
		\hline 
    	\hline 
		Whisper~\cite{whisper} &Speech to Text (S2T) & MIT   \\   
            &  Language Detection (LD) &\\
            \hline
		mBart~\cite{mbart_model} &English Translation (ET) &MIT            \\
		\hline 
		PANN~\cite{kong_pretrain} &Acoustic Event Detection (AED)  &MIT     \\    
		    	\hline 
        pyannote~\cite{pyannote}& Speaker Count (SC) &  MIT \\ \hline
     	ECAPA-TDNN~\cite{ecapa} &Voice Gender Detection (VGD) & MIT\\
     	    	    	\hline      	
     SpeechBrain~\cite{speechbrain}	&Speaker Diarization (SD) &     Apache license 2.0   \\
        \hline
            Wav2vec2~\cite{speechbrain, wav2vec2}
          &Speech Emotion Recognition (SER) & Apache license 2.0\\ 
     	    	    	\hline 

            KD~\cite{ac} &Audio Captioning (AC) & Apache license 2.0\\
     	    	    	\hline 
AIT-ASC~\cite{lampham_01, lampham_02} &Acoustic Scene Classification (ASC) & \textbf{Our development}     \\        
     	    	\hline 
            AIT-DSD~\cite{lam_pp_01, lam_pp_02, lam_group_01} & Deepfake Speech Detection (DSD) & \textbf{Our development}     \\        
     	    	\hline                          
        \end{tabular}  
	}  
	\label{table:model} 
\end{table}
\begin{figure*}[t]
    \centering
    \includegraphics[width=0.9\linewidth]{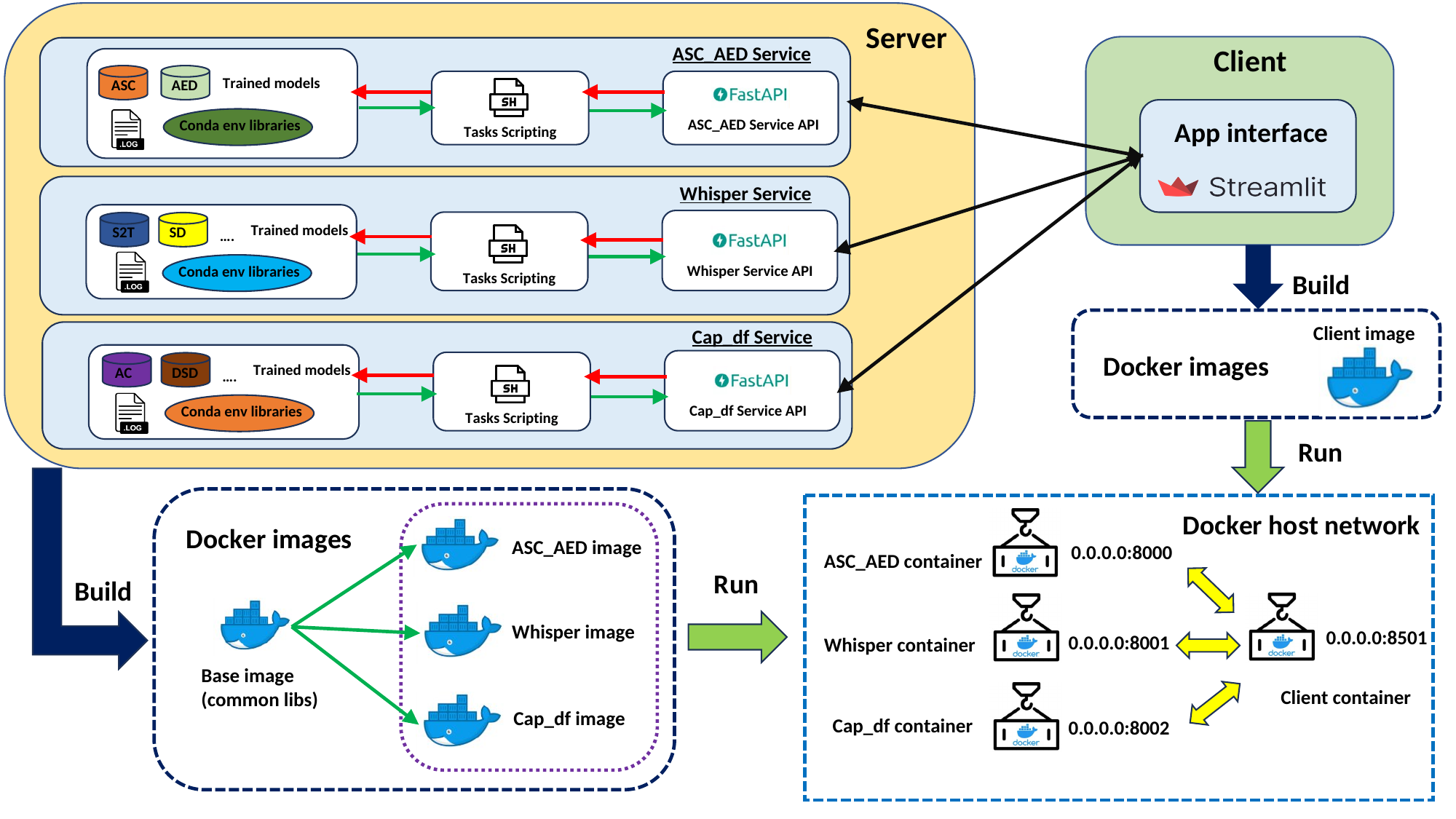}
    \caption{Deployment of our system as multiple services leveraging Docker}
    \label{docker_deploy}
\end{figure*}
The final group is referred to as summary-based tasks which provide a high-level understanding of the audio content by summarizing, assessing its overall characteristics. 
They include Audio Captioning (AC)~\cite{ac} (e.g., generating textual descriptions summarizing the content of the audio recording) and Speech Deepfake Detection (DSD)~\cite{lam_pp_01, lam_pp_02} (e.g., determining whether a given speech is artificially manipulated or normal real speech). As illustrated in Fig.~\ref{fig:audio_logic}d, models in this task group process input audio based on fixed-length segments. The Audio Captioning (AC) model operates on 10-second segments to accommodate potential shifts in the overall content of the recording, while the DSD model processes 4-second segments, which are sufficient to capture distinctive artifacts in human speech. The final output of the DSD model is obtained through majority voting on the segment-wise results. Again, the outputs from this task group are then gathered and collectively referred to as `Other information'.

Given JSON file results from task groups, these JSON files are then aggregated and standardized.
The Standardize module aims to re-structure the final information of the audio recordings. In our work, we employ a timeline-centric standardization approach that 
arranges the results based on a shared timeline, where single task's outputs are aligned according to their occurrence throughout the audio recording.

By structuring the AAP component into such content-based categories, our proposed tool ensures a systematic and scalable approach to extracting meaningful insights from recordings. This modular framework allows for greater expansion and management (i.e., new tasks or improved models can be categorized and integrated to group based on their result's contents without re-designing the entire system). 

\subsection{Audio Retrieval Phase (ARP)}
The Audio Retrieval Phase (ARP) plays a crucial role in bridging the gap between audio analysis results from the AAP and a human-interpretable response. Given the final information JSON output from the AAP component, which may be fragmented or lack context, the ARP involves leveraging a Large Language Model (LLM) model to processes and contextualizes these information and then transfer them into a coherent and informative response tailored to the user’s query. 
Since the application is designed for surveillance purposes, the ARP needs to address several challenges such as irrelevant queries/questions (e.g., Users ask about topics unrelated to the audio content, leading to off-topic or unhelpful responses), LLM Hallucinations (e.g., the LLM might generate inaccurate or misleading information despite being provided the extract audio content), etc. To ensure reliable and contextually relevant responses, we explore two prompting techniques:

\textbf{Zero-shot Prompting with Filtering \& Response Control (ZPFR):}  Zero-shot prompting is a technique in which an AI model is given a task or question without any prior examples or specific training, relying solely on its pre-existing knowledge and prompt to generate a response~\cite{zeroshot}. While effective in many cases, zero-shot prompting alone may not guarantee precision or adherence to specific guidelines. To improve response quality, we integrate filtering rules into the system prompt to pre-process user queries and eliminate irrelevant inputs. Additionally, response control rules are applied to guide the model in generating outputs that align with pre-defined objectives. An example of such type of prompting is illustrated in Fig.~\ref{fig:zpfr}.


\textbf{Chain-of-Thought Prompting:} Chain-of-thought (CoT) prompting~\cite{cot} is a technique that significantly improves the reasoning capabilities of large language models (LLMs). Instead of simply asking for a final answer, CoT prompting encourages the model to break down complex problems into a series of intermediate steps, mimicking human-like reasoning. Based on this mechanism, we try to guide the LLM toward various reasoning steps to implicitly handling irrelevant queries and reducing hallucination above before generating final responses. An example of this type of prompt can be described as Fig.~\ref{fig:cot}.


\section{Deployment}

\subsection{Module Implementation}

To deploy the proposed Aud-Sur tool, we first implement the User Interface (UI), independent audio models used in the AAP component, and LLM in the ARP component.
The first component of User Interface (UI) is implemented by Streamlit platform.
Regarding audio models in the AAP component, the Table~\ref{table:model} presents the pre-trained audio models along with their respective licenses for  specific audio tasks.
All audio models leveraged deep neural networks and were implemented with Pytorch framework. 
As we propose our self-developing models (Acoustic Scene Classification and Deepfake Audio Detection) and leverage open-source models for other audio tasks, it is feasible to further expand the scope of the AAP component and share for the research community. Regarding the ARP component, we use two approaches to leverage the operation of Large Language Model (LLM). The first approach involves interacting with the LLM via an API, while in the second one, we runs the LLM locally, providing greater control and data privacy. In both cases, we utilize LLaMA~\cite{lama} as our LLM with the following configurations: a temperature of 0.3, a maximum token limit of 256 to constrain responses, and a top-p value of 0.5 for nucleus sampling.

\subsection{Deployment Environment}
\begin{figure}[t]
    \centering
    \includegraphics[width=1.0\linewidth, height = 0.6\linewidth]{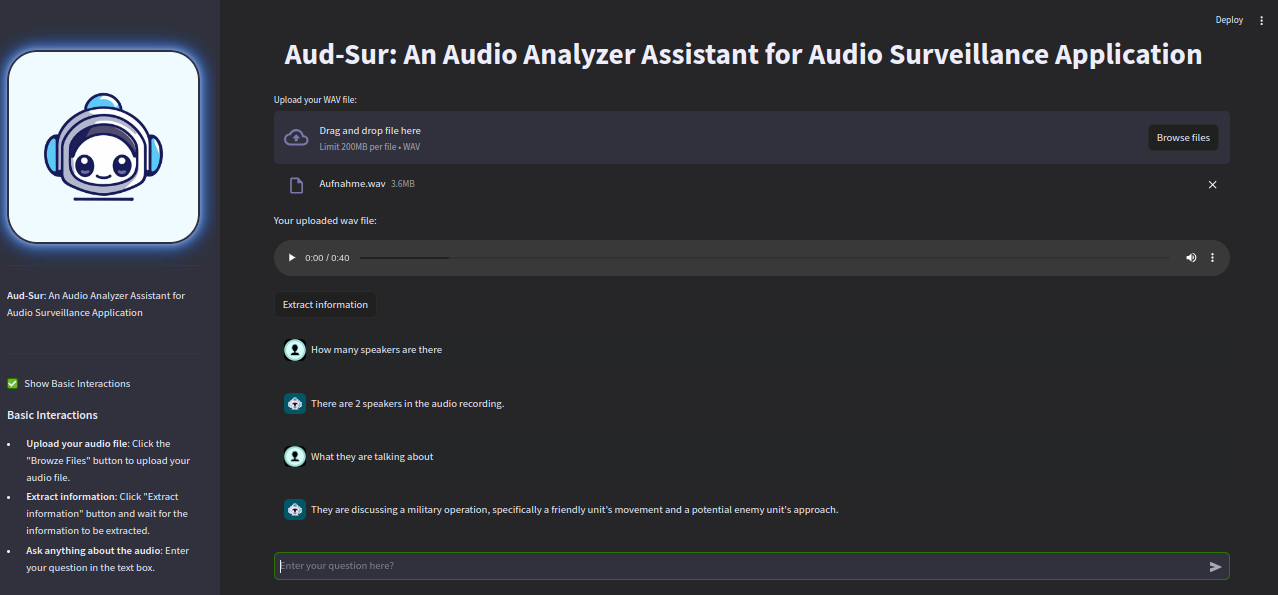}
    \caption{The user interface (UI) of our proposed Aud-Sur tool}
    \label{interface}
\end{figure}
We utilize Docker and FastAPI to deploy our proposed Aud-Sur tool following a client-server architecture as illustrated in Fig.~\ref{docker_deploy}. 
The client side consists of a user interface (UI). This UI allows users to interact with the system and submit queries. The queries and response results are processed with the support from prompting designs and a Large Language Model (LLM) operating. 
The server side is structured as multiple independent services, each of service is responsible for a predefined group of tasks as mentioned in section~\ref{architecture}. 
In other words, each service includes some audio components which use the same Python package libraries and Conda dependencies. 
Each service also includes independent API which is used to communicate via predefined ports and collectively generate results for the client side.
Regarding Docker containerization, described as the bottom of Fig~\ref{docker_deploy}, we first construct a base Docker image that includes common dependencies (e.g., Conda environments, GPU-enabled inference installations, and essential libraries). 
Building upon the based image, each service is packaged into a separate Docker image and run independently in its own Docker container. 
The services interact via API calls and results are transferred to the client through port connections. 
By following this approach, we achieve a modular, scalable and efficient deployment process, ensuring independent execution of all system functionalities.

\begin{figure}[t]
    \centering
    \includegraphics[width=0.95\linewidth, height = 0.6\linewidth]{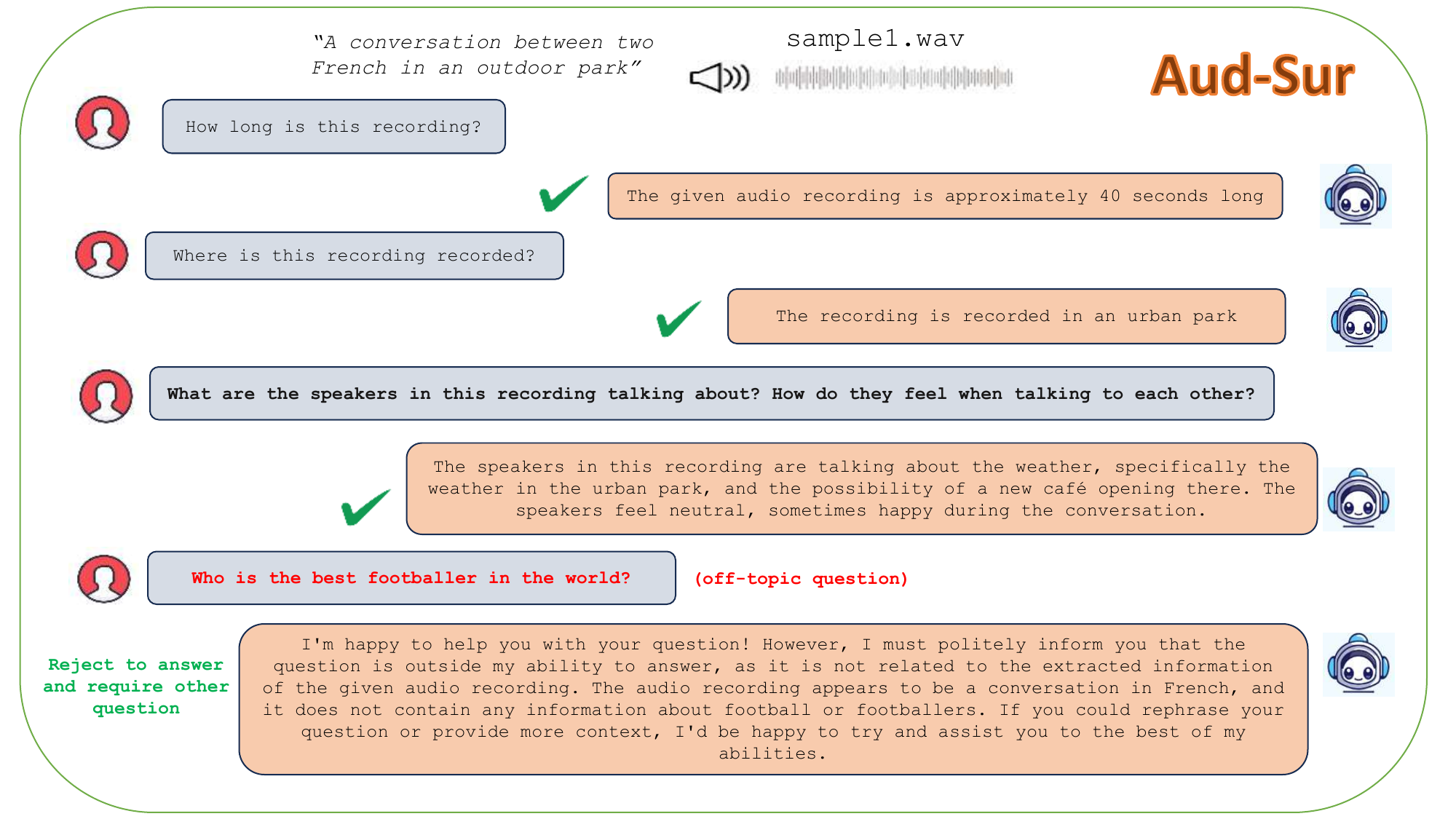}
    \caption{Our proposed Aud-Sur tool operates as a interactive assistant}
    \label{demo-conversation}
\end{figure}

\section{Demonstration and Discussion}
\label{demo}
\begin{figure*}[th]
    \centering
    \includegraphics[width=1.0\linewidth]{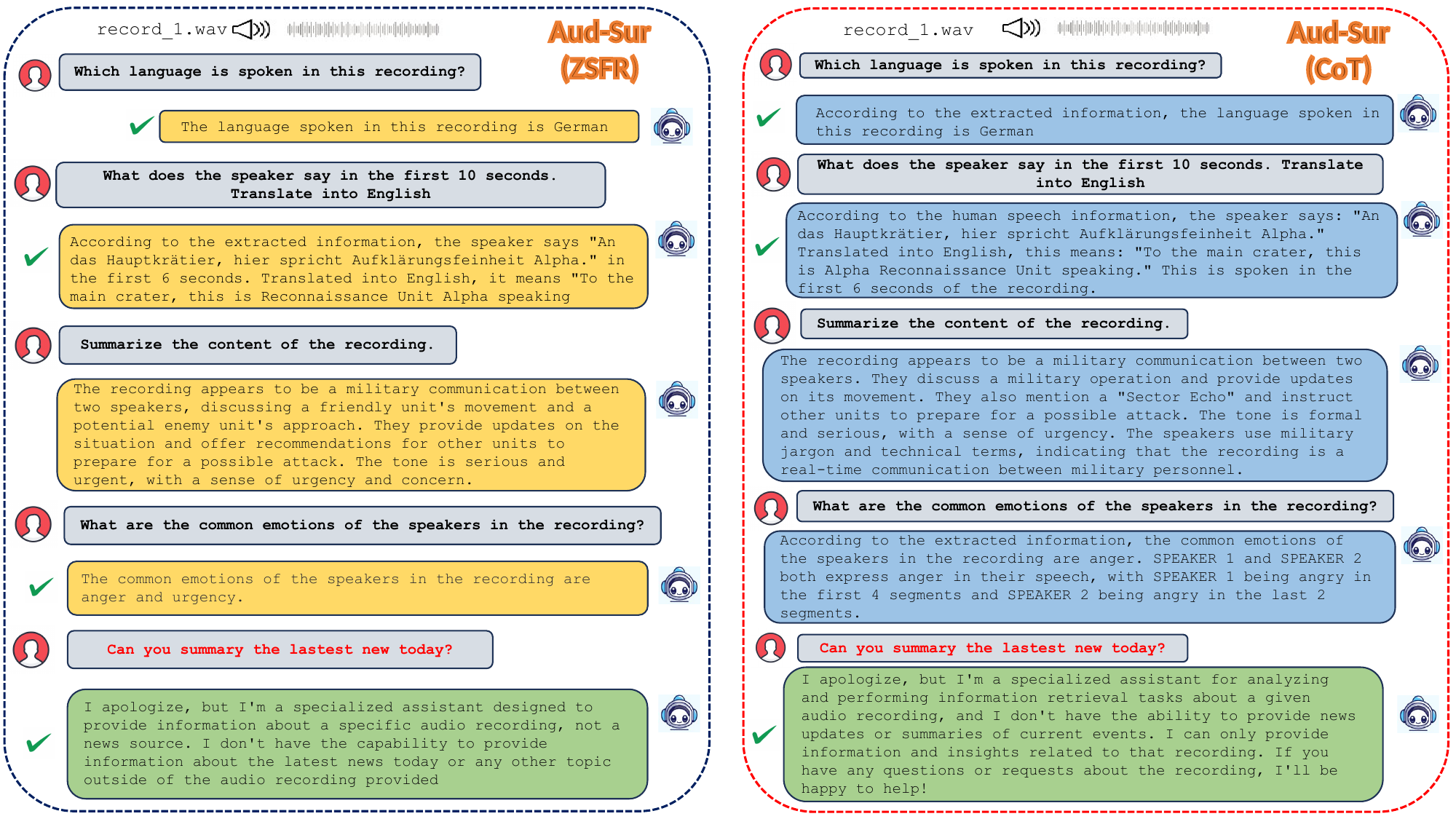}
    \caption{Compare LLM response using ZPFR prompting and CoT promting (record\_1.wav involves a confidential reconnaissance unit report between two German speakers)}
    \label{fig:sample1}
    \vspace{6 mm}
\end{figure*}

\begin{figure*}[th]
    \centering
    \includegraphics[width=1.0\linewidth]{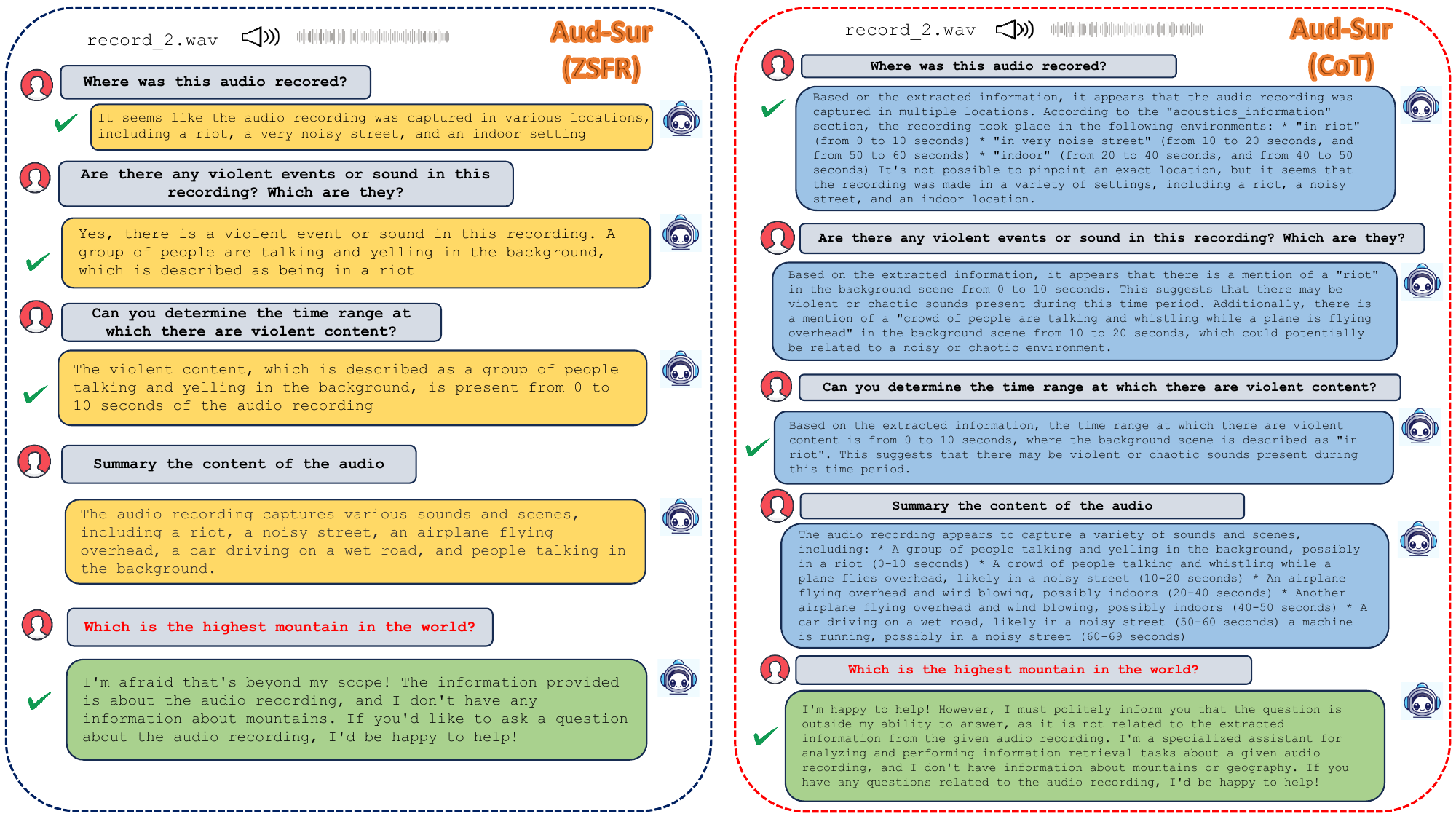}
    \caption{Compare LLM response using ZPFR prompting and CoT promting (record\_2.wav file is created by combining three completely audio recordings, with one scene of riot context at the first 15 seconds.)}
    \label{fig:sample2}
        \vspace{6 mm}

\end{figure*}
This subsection presents a live demonstration of our proposed Aud-Sur tool. Fig.~\ref{interface} illustrates the user interface, where users can upload an audio file in .wav or .mp3 format and submit text-based queries or questions to inquire information related to the uploaded audio. The tool processes each query/question by analyzing the input, extracting relevant information, and then generating a corresponding response. The interaction between the user and the system is depicted in Fig.~\ref{demo-conversation}. 
Unlike conventional methods that rely on a single model’s input-output pipeline, the integration of a LLM in the ARP component of our proposed Aud-Sur tool enables the tool function in a chatbot-like manner that facilitates a more dynamic and context-aware approach to information retrieval. This approach overcomes the limitations of standalone audio models, particularly in cases where a user's query requires information from multiple models or consists of multiple sub-questions. For instance, in Figure~\ref{demo-conversation}, the user's third query "What are the speakers in this recording talking about? How do they feel when talking to each other?" requires results from both speech-to-text and speech emotion recognition. Leveraging multiple-model approach combined with the support of LLM, our Aud-Sur tool could combines information from multiple sources to generate a comprehensive response. Additionally, with prompting support, our system can filter out queries that are irrelevant to the audio content. As illustrated in Fig~\ref{demo-conversation}, the system politely rejects off-topic questions (highlighted in red) and instead prompts the user with more contextually appropriate questions.

To assess the impact of different prompting techniques, we evaluate our tool’s responses using ZPFR and CoT prompting on the same set of user queries. 
Given that our tool is designed for surveillance applications, we specifically analyze its ability to monitor and detect audio indicators related to potential threats, such as extreme audio events, riots, violent sounds, pre-crime activities, etc. For these purposes, we conduct demonstrations using two audio samples that contain diverse information relevant to our objectives. 
In particular, the first samples named as `record\_1.wav` features a confidential reconnaissance unit report conversation between two German speakers, aiming at evaluating our system’s ability to extract sensitive intelligence-related information in human speech.
The second one, referred to as `record\_2.wav`, is created by combining three different audio recording with different background scenes, with a riot occurring in the first 15 seconds. The second recording is designed to test the system’s effectiveness in detecting security threats and handle context shifting.

As shown in Fig.~\ref{fig:sample1} and Fig.~\ref{fig:sample2}, our Aud-Sur tool successfully extracts relevant information in both cases, demonstrating its robustness across varied audio environments. The analysis also shows differences in the results using the two prompting techniques. In particular,  ZPFR method tends to provide more natural, controlled and compact responses. In contrast, CoT prompting produces more structured and detail-oriented, due to its predefined reasoning steps. These differences highlight the importance of optimizing a prompting strategy to enhance accuracy, reliability, and suitability of LLM-driven real-world audio surveillance applications.


\section{Conclusion}
We have presented Aud-Sur, an audio analyzer assistant tool for audio surveillance applications, by leveraging various audio models for audio information extraction and incorporates a large language model (LLM) for natural language-based audio information retrieval.
As multiple open-source and audio models are integrated into the Aud-Sur tool and a microservice-based architecture with Docker is used for the deployment process, our system enables seamless integration of new audio model/tasks, making the tool easily adaptable and widely shared for various audio surveillance applications.

\section*{ACKNOWLEDGMENTS}
EUCINF project is co-funded by European Union under grant agreement N°101121418.
Views and opinions expressed are however those of the author(s) only and do not necessarily reflect those of the European Union or the European Commission. Neither the European Union nor the granting authority can be held responsible for them.

Defame Fakes is funded by the Austrian security research programme KIRAS of the Federal Ministry of Finance (BMF).

\begin{figure}[h]
    \centering
    \includegraphics[width =1.0\linewidth]{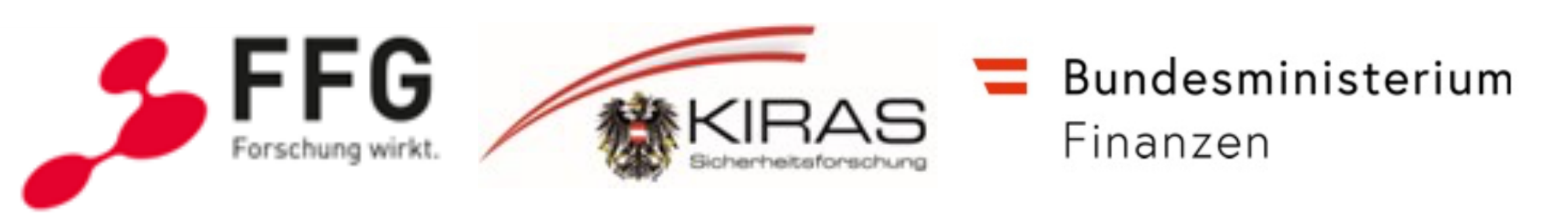}
       	\vspace{-0.4cm}
\end{figure}

\end{document}